%% file: WFCE.tex
\documentclass[aps,prb,twocolumn,superscriptaddress,floatfix]{revtex4-1}

\usepackage{bm}
\usepackage{amssymb}
\usepackage{graphicx}
\usepackage{amsmath}
\usepackage{dcolumn}
\usepackage[pdftex]{epsfig}
\usepackage{color}
\usepackage[german,english]{babel}
\usepackage{psfrag}
\usepackage{hyperref}
\usepackage[caption=false]{subfig}
\usepackage{longtable}
\usepackage{pstricks}
\usepackage{pst-node}
\usepackage[utf8]{inputenc}

\usepackage{verbatim}

\newcommand{\mv}[1]{\mathbf{#1}}
\newcommand{\mychi}{\protect\raisebox{2pt}{$\chi$}}

\begin{document}

\title{Wannier Function Approach to Realistic Coulomb Interactions in Layered Materials and Heterostructures}

\author{M. R\"osner}
\affiliation{
  Institut f{\"u}r Theoretische Physik, 
  Universit{\"a}t Bremen, 
  Otto-Hahn-Allee 1, 
  D-28359 Bremen, Germany
  }
\affiliation{
  Bremen Center for Computational Materials Science, 
  Universit{\"a}t Bremen, 
  Am Fallturm 1a, 
  D-28359 Bremen, Germany
  }

\author{E. \c{S}a\c{s}{\i}o\u{g}lu}
\author{C. Friedrich}
\author{S. Bl\"{u}gel}
\affiliation{
  Peter Gr{\"u}nberg Institut and Institute for Advanced Simulation, 
  Forschungszentrum J{\"u}lich and JARA, 
  D-52425 J{\"u}lich, Germany
  }
\author{T. O. Wehling}
\affiliation{
  Institut f{\"u}r Theoretische Physik, 
  Universit{\"a}t Bremen, 
  Otto-Hahn-Allee 1, 
  D-28359 Bremen, Germany
  }
\affiliation{
  Bremen Center for Computational Materials Science, 
  Universit{\"a}t Bremen, 
  Am Fallturm 1a, 
  D-28359 Bremen, Germany
  }

\date{\today}

\begin{abstract}
We introduce an approach to derive realistic Coulomb interaction terms in free standing layered materials and vertical heterostructures from ab-initio modelling of the corresponding bulk materials. To this end, we establish a combination of calculations within the framework of the constrained random phase approximation, Wannier function representation of Coulomb matrix elements within some low energy Hilbert space and continuum medium electrostatics, which we call Wannier function continuum electrostatics (WFCE). For monolayer and bilayer graphene we reproduce full ab-initio calculations of the Coulomb matrix elements within an accuracy of $0.2$~eV or better. We show that realistic Coulomb interactions in bilayer graphene can be manipulated on the eV scale by different dielectric and metallic environments. A comparison to electronic phase diagrams  derived in [M. M. Scherer et al., Phys. Rev. B \textbf{85}, 235408 (2012)] suggests that the electronic ground state of bilayer graphene is a layered 
antiferromagnet and remains surprisingly unaffected by these strong changes 
in the Coulomb interaction.
\end{abstract}


\maketitle

\section{Introduction}
Since the isolation of monolayer graphene\cite{novoselov_electric_2004}, the library of experimentally available two dimensional (2D) materials has been continuously growing\cite{novoselov_two-dimensional_2005,nicolosi_liquid_2013} and includes now e.g. graphene, hexagonal boron nitride, silicene or the class of transition metal dichalcogenides (TDMCs). Together with recent developments in on-demand stacking of these atomically thin crystals, a whole new class of hybrid materials is coming into reach, now\cite{geim_van_2013}. While these 2D structures hold promises from the realization of exotic electronic quantum phases at elevated temperatures\cite{fogler_high-temperature_2014,rodin_excitonic_2013} to concepts for optoelectronic information processing and light harvesting \cite{wang_electronics_2012,britnell_strong_2013}, their theoretical description is very challenging as they generally combine structural complexity with pronounced electronic interaction effects such as renormalized quasiparticles or 
competing electronic phases \cite{kotov_electron-electron_2012-1}.

One general theoretical strategy towards realistic descriptions of interacting electron systems is the combination of ab-initio and model Hamiltonian approaches, where single particle and interaction parameters are derived from first principles calculations\cite{pavarini_lda+dmft_2011}. Here, one typically considers quantum lattice models like extended multiband Hubbard models, which describe electrons within some set of low energy bands interacting with each other. Realistic Coulomb interaction matrix elements entering these models should be appropriately screened, i.e. account for screening due to those states which are not exciplicitly treated in the low energy models, and can be calculated from first principles using the so-called constrained Random Phase Approximation (cRPA) \cite{aryasetiawan_frequency-dependent_2004}. The computational demand of these calculations is comparable to $GW$ calculations, which makes the treatment of complex heterostructures e.g. in plane wave based approaches very 
challenging.

Essentially two different strategies have been developed to circumvent computational problems in obtaining appropriately screened interactions for thin films or layered materials. First, in long-wavelength approaches to layered materials model dielectric functions based on a description of the two-dimensional screening in terms of macroscopic electrodynamics can be straightforwardly employed (see e.g. Refs. \onlinecite{MacDonald2010,Katsnelson_drag_2011}). Second, modified Coulomb interactions involving e.g. a truncation in the vertical direction\cite{Thygesen_PRB2013} or unscreening in terms of model dielectric functions\cite{Scheffler_Schindlmayr_PRB08} can be employed directly on the fully microscopic $GW$ level to reach faster convergence of the screened interactions in repeated slab approaches. While the first set of approaches comes at the advantage of almost no computation cost in obtaining screened Coulomb interactions, it generally relies on a-priori unknown adjustable parameters. The second set of 
approaches contains all microscopic real material informations but requires a new fully microscopic calculation when the dielectric environment of some layered material (e.g. the substrate) is changed and remains still computationally demanding.

In this paper, we introduce a bridge between these two complementary classes of approaches and develop an approximate very simple yet accurate approach to derive realistic Coulomb interaction matrix elements for electrons in free standing layered materials and vertical heterostructures from cRPA modelling of the corresponding bulk materials. To this end, we combine Wannier function representations of the Coulomb matrix elements within some low energy Hilbert space of interest with continuum medium electrostatics, as we explain in section \ref{sec:WFCE}. This allows us to avoid repeated slab calculations on the cRPA level. In section \ref{sec:graphite_to_graphene}, we illustrate our Wannier function based approach with the example of graphene, bilayer graphene as well as related heterostructures like Ir intercalated graphene. A particular advantage of the approach introduced, here, is that one can very easily assess how different environments affect Coulomb interactions in realistic layered materials, as we 
show with the example of bilayer graphene in section \ref{sec:bilayer}.

\section{Model Hamiltonian and the cRPA approach\label{sec:abinitio}}
For the graphene based systems to be studied, here, we consider an effective model, which includes the carbon $p_z$ orbitals (i.e. the $\pi$ bands) and treats the $\sigma$ bands as well as states at higher energies as the "rest"\cite{wehling_strength_2011}. We can describe such a system with a generalized Hubbard model for the $p_z$ orbitals with the many-body Hamiltonian
  \begin{align} \label{eqn:Hubbard}
    H = &-t \sum_{<\mv{i},\mv{j}>, \sigma} c_{\mv{i}\sigma}^\dagger c_{\mv{j}\sigma} 
        + U_{\mv{00}} \sum_\mv{i} n_{\mv{i}\uparrow} n_{\mv{j}\downarrow} \\
        &+ \frac{1}{2} \sum_{\substack{\mv{i}\neq\mv{j} \\ \sigma,\sigma'}} U_{\mv{i}\mv{j}} n_{\mv{i}\sigma} n_{\mv{j}\sigma'}, \nonumber
  \end{align}
where $c_{\mv{i}\sigma}$ annihilates an electron with spin $\sigma \in \{ \uparrow, \downarrow \}$ at site $\mv{i}$ and $n_{\mv{i}\sigma} = c_{\mv{i}\sigma}^\dagger c_{\mv{i}\sigma}$. The index $\mv{i} = (i, \text{A or B})$ labels the sublattice (A, B) and the unit cell centered at position $\mv{R}_i$.

The cRPA\cite{aryasetiawan_frequency-dependent_2004} is used to derive the effective partially screened Coulomb interaction terms $U_{\mv{i}\mv{j}}$ entering the model of Eq. (\ref{eqn:Hubbard}) from first principles. Within the standard random phase approximation (RPA) the screened Coulomb interaction is defined as $W(\mv{q},\omega) = [1 - v(\mv{q}) P(\mv{q},\omega)]^{-1}v(\mv{q})$ 
with the bare Coulomb repulsion $v(\mv{q})$ and the momentum and frequency dependent
polarization function $P(\mv{q},\omega)$. We note that all quantities are non-local
in space, e.g., $v(\mv{r},\mv{r}',\mv{q})$, where $\mv{r}$ and $\mv{r}'$ are restricted to the unit
cell. Using a suitable basis the functions become matrices and the equation above becomes a matrix equation with $1$ being the unit matrix.

To obtain the effective Coulomb interaction, the function $P(\mv{q}, \omega)$ is divided into a $P_\pi$ and a $P_r$ part. The former accounts for charge fluctuations that are induced by virtual transitions within the $\pi$ bands only, i.e., from $\pi$ to $\pi^*$. All other transitions are included in the remainder $P_r$. The latter give rise to screening effects that reduce the effective Coulomb interaction between the $\pi$ electrons of the model
  \begin{align}
    U(\mv{q}, \omega) = \frac{v(\mv{q})}{1 - v(\mv{q}) P_r(\mv{q}, \omega)} = \epsilon^{-1}(\mv{q}, \omega) v(\mv{q}),
    \label{eqn:screenedU}
  \end{align}
  where $\epsilon(\mv{q}, \omega)$ is the matrix representation of the microscopic
dielectric function $\epsilon(\mv{r},\mv{r}',\mv{q},\omega)$ and $\epsilon^{-1}$ its inverse.
  The interaction $U$ is in general momentum and frequency dependent and can be long ranged.
  
  We evaluate these bare and screened Coulomb matrix elements in an all-electron mixed product basis\cite{friedrich_efficient_2009,friedrich_efficient_2010} based on the full-potential linearized augmented-plane-wave (FLAPW) method \cite{weinert_total-energy_1982, koelling_use_1975, andersen_linear_1975}. For this purpose we start with a density functional theory (DFT) calculation employing the FLEUR code \cite{_juelich_2014, grotendorst_computational_2006} to obtain the corresponding ground states within the FLAPW method. Afterwards, we use the SPEX code \cite{friedrich_efficient_2009, friedrich_efficient_2010} to calculate the bare and screened Coulomb matrix elements in the constrained random phase approximation \cite{sasioglu_effective_2011}. Technical details concerning these calculations can be found in the appendix \ref{sec:appendix_abinitio}.

\section{Combination of Wannier Functions and Continuum Electrostatics}\label{sec:WFCE}
In this section, we explain our approach to derive appropriately screened Coulomb interaction matrix elements for electrons in two dimensional materials (e.g. graphene) and their heterostructures on the basis of Coulomb interaction matrix elements from parent three dimensional bulk systems (e.g. graphite). To this end,
 we recall first the continuum electrodynamic description of layered materials, free standing monolayers as well as heterostructures. Afterwards, the continuum formulation is embedded into a quantum lattice description in terms of localized Wannier functions.
  
  \subsection{Continuum electrostatic description}\label{sec:cont_el}
  
    \begin{figure}[ht!]
      \begin{center}
	\includegraphics[width=0.95\linewidth]{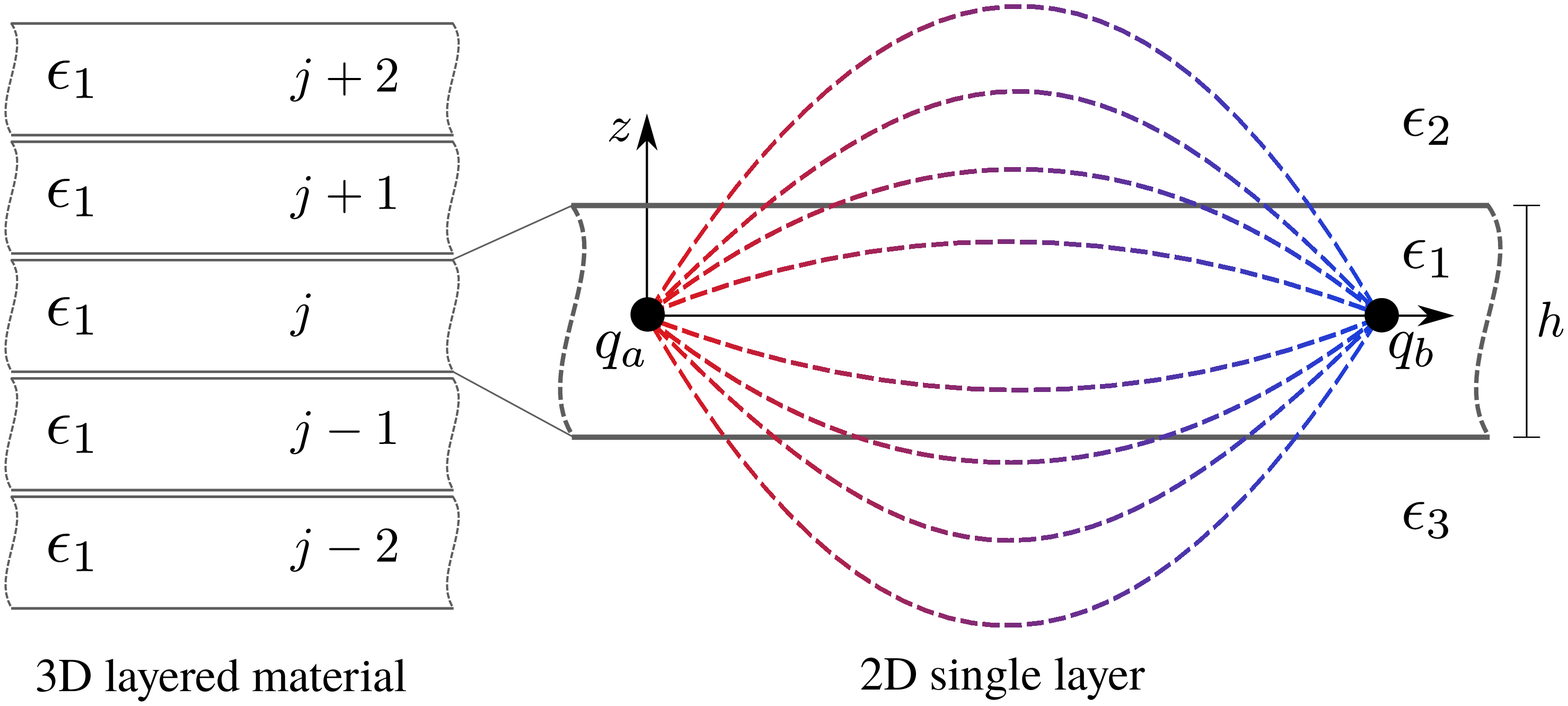}
	\caption{(Color online) \textbf{Left:} Three dimensional layered material. The infinite stack of single layers is numbered by $j$. \textbf{Right:} Corresponding two dimensional monolayer in a variable dielectric surrounding. The arising electric field of two charges $q_a$ and $q_b$ is indicated by dashed lines. The dielectric properties in the layer is denoted by $\varepsilon_1$, which is actually a non-local and thus $\mv{q}_\parallel$ dependent function (see main text). \label{fig:DielectricModel}}
      \end{center}
    \end{figure}
    
    The problem of screening in terms of continuum electrostatics in a layered material is illustrated in Fig. \ref{fig:DielectricModel} and has been considered in Refs. \onlinecite{keldysh_l.v._coulomb_1979, jena_enhancement_2007, emelyanenko_effect_2008}. We generalize these works to include \emph{non-local} screening effects. To this end, we start with the Poisson equation in the presence of matter:
    \begin{align}
      \nabla \int d^3\mv{r'} \left[ \delta(\mv{r}-\mv{r'}) + \mychi(\mv{r}-\mv{r'}) \right]  \mv{E}(\mv{r'}) = \frac{1}{\varepsilon_0} \rho_\text{ext}(\mv{r}),
    \end{align}
 which relates the divergence of the dielectric displacement (left) to
the density of the external, i.e., unbound, charge (right). The
dielectric displacement is written in terms of the electric field $\mv{E}(\mv{r'})$
and the induced polarization $\int \mychi(\mv{r}-\mv{r'}) \mv{E}(\mv{r'}) d^3r'$, where the
susceptibility $\mychi$ is generally non-local but depends only on differences of the spatial coordinates $\mv{r}-\mv{r}'$.  

These coordinates can be separated into the in-plane and vertical components according to $\mv{r} = ( \mv{r}_\parallel, z)$:
    \begin{align}
      \nabla \int &d^3\mv{r'} \left[ \delta(\mv{r}_\parallel-\mv{r'}_\parallel) \delta(z-z') + \mychi(\mv{r}_\parallel-\mv{r'}_\parallel, z-z') \right] \label{eq:Poisson_bulk}\\ \notag
      & \cdot \mv{E}(\mv{r'}_\parallel, z') = \frac{1}{\varepsilon_0} \rho_\text{ext}(\mv{r}_\parallel, z).
    \end{align}
    The electric field $\mv{E}(\mv{r'}_\parallel, z')$ can be obtained from the electrostatic potential  $\mv{E}(\mv{r'}_\parallel, z')=-\nabla \Phi(\mv{r'}_\parallel, z')$. 
    
 Due to the translational invariance of the continuum medium, Eq. (\ref{eq:Poisson_bulk}) becomes diagonal in
Fourier space and shows a simple algebraic relation between the susceptibility, the potential, and the external charge density 
    \begin{align}
      q^2 \left[ 1 + \mychi(\mv{q}_\parallel, q_z) \right] \Phi(\mv{q}_\parallel, q_z) &= \frac{1}{\varepsilon_0} \rho_\text{ext}(\mv{q}_\parallel, q_z). \label{eq:Poisson_bulk_fourier}
    \end{align}
    Instead of the susceptibility we can consider the dielectric function  $\varepsilon(\mv{q}_\parallel,q_z) =  1 + \mychi(\mv{q}_\parallel, q_z)$ and express the solution of the Poisson equation (\ref{eq:Poisson_bulk_fourier}) in the form 
    \begin{equation}
      \Phi(\mv{q}_\parallel, q_z)=\frac{1}{\varepsilon_0 \varepsilon(\mv{q}_\parallel,q_z) q^2}\rho_\text{ext}(\mv{q}_\parallel, q_z)=\frac{\Phi_\text{ext}(\mv{q}_\parallel, q_z)}{\varepsilon(\mv{q}_\parallel,q_z)}, \label{eq:phi_bulk}
    \end{equation}
    where we introduced $\Phi_\text{ext}(\mv{q}_\parallel, q_z)$, which is the electrostatic potential created by
the external charge.

In turn, Eq. (\ref{eq:phi_bulk}) can be regarded as a definition of the dielectric
function: \begin{align}
      \varepsilon(\mv{q}_\parallel, q_z) = \frac{\Phi_\text{ext}(\mv{q}_\parallel, q_z)}{\Phi(\mv{q}_\parallel, q_z)}. \label{eqn:epsEff}
    \end{align}
On the other hand, the dielectric function can be obtained
from first principles, for example using the random-phase approximation.
In this context, $\Phi_\text{ext}(\mv{q}_\parallel, q_z)$ plays the role of the bare Coulomb potential, which is responsible for the bare Coulomb interaction $v^\text{3D}(\mv{q}_\parallel, q_z)=-e\Phi_\text{ext}(\mv{q}_\parallel, q_z)$ and $\Phi(\mv{q}_\parallel, q_z)$ comprises $v^\text{3D}(\mv{q}_\parallel, q_z)$ and the potential created by the induced
charge. So, $\Phi(\mv{q}_\parallel, q_z)$ corresponds to the screened interaction $U^\text{3D}(\mv{q}_\parallel, q_z)$.

We now proceed to relate the bulk dielectric function to the dielectric
function of a two-dimensional system embedded into a dielectric
environment, as shown in Fig. \ref{fig:DielectricModel}. We assume that the former has been determined from first
principles. As we will explicate later, we only modify the leading
eigenvalue of the microscopic dielectric function
$\varepsilon(\mv{r},\mv{r}',\mv{q}_\parallel,q_z)$ so that we may assume $\varepsilon(\mv{q}_\parallel,q_z)$ to
be a scalar function. The embedding is not as easily done as in Refs.
\onlinecite{emelyanenko_effect_2008, jena_enhancement_2007, keldysh_l.v._coulomb_1979}, because we face the problem of non-localities, in particular the
$\mv{q}_z$ dependence, which describes the periodicity of the bulk material in $z$-direction. Clearly, an assumption on how non-localities translate from bulk to monolayer or heterostructures has to be made. Here, we continue with the simplest possible approximation and neglect all non-localities in $z$-direction, i.e. we replace
    \begin{equation}
      \varepsilon(\mv{q}_\parallel, q_z) \rightarrow \varepsilon_1(\mv{q}_\parallel)= \frac{h}{2\pi} \int_{-\pi/h}^{\pi/h} {\rm d}q_z\varepsilon(\mv{q}_\parallel,q_z), \label{eqn:eps_z_local}
    \end{equation}
    where $h$ plays the role of an effective layer thickness. This definition is plausible as the
two-dimensional embedding breaks the periodicity in $z$-direction, and
only the local term of $\varepsilon$ should remain relevant. In this sense, Eq. (\ref{eqn:eps_z_local}) gives this local
term as the Fourier transformation to the center of the monolayer, i.e., $z=0$. A similar formula was used in Ref. \onlinecite{Thygesen_PRB2013} to define a "two-dimensional macroscopic dielectric function".

We now assume that $\varepsilon_1(\mv{q}_\parallel)$ is constant on the whole width of the monolayer, i.e., for $|z|\le h/2$, and
consider two dielectric materials on both sides with dielectric constants $\epsilon_2$ and $\epsilon_3$ according to Fig. \ref{fig:DielectricModel}, which gives
    \begin{align}
      \varepsilon(\mv{q}_\parallel, z) = \left\{
	  \begin{array}{lccc} 
	    \varepsilon_2 \qquad & 		z &> &\frac{h}{2} \\
	    \varepsilon_1(\mv{q}_\parallel) & |z| &\le  &\frac{h}{2} \\
	    \varepsilon_3 \qquad & 		z &< &\frac{-h}{2} \\
	  \end{array}\right.. \label{eq:eps_film}
    \end{align}

To find the appropriately screened interaction $U^\text{2D}(\mv{q}_\parallel)$ between two electrons in the central layer, we consider the electrostatic problem of an oscillating two dimensional charge density $\rho(\mv{q}_\parallel,z)=\rho^{2D}(\mv{q}_\parallel)\delta(z)$ in the center of the monolayer, which is at $z=0$. The resulting electrostatic potential $\Phi(\mv{q}_\parallel, z)$ is not the same as in the bulk [Eq. (\ref{eq:phi_bulk})] due to modified screening and the fact that we are now considering a 2D charge density confined to $z=0$. 

In the dielectric continuum model, the modification of the screening is caused by the formation of image charges at the interfaces between the dielectrics. Let us first consider a single interface, where the dielectric constant changes from $\epsilon_1$ to $\epsilon_2$, and a test charge $q$ in region $1$. For an observer in region $1$, the induced polarization has the form of an image charge of magnitude $q(\epsilon_1-\epsilon_2)/(\epsilon_1+\epsilon_2)$ that is located in region $2$ at an equal distance from the interface as the test charge \cite{JackClas1999}. If one has more than one interface, as in the case of our dielectric model, the charge is reflected infinitely many times (as light between two parallel optical mirrors), giving rise to an infinite number of image charges of ever decreasing magnitude perpendicular to the interfaces. In the more general case of the two-dimensional charge distribution, each Fourier component of $\rho(\mv{r})$ is mirrored according to the corresponding component of the 
dielectric function and the ratio is
    \begin{align}
      \tilde{\varepsilon}_j(\mv{q}_\parallel) = \frac{\varepsilon_1(\mv{q}_\parallel) - \varepsilon_j}{\varepsilon_1(\mv{q}_\parallel) + \varepsilon_j}
    \end{align}
with $j=2$ and $j=3$ for the upper and lower interface, respectively. Several works have treated this situation, a dielectric monolayer sandwiched between two semi-infinite dielectric materials \cite{keldysh_l.v._coulomb_1979, jena_enhancement_2007, emelyanenko_effect_2008}. We use here a special case of a formula derived in Ref. \onlinecite{emelyanenko_effect_2008}, giving the effective two-dimensional dielectric function
\begin{widetext}
    \begin{equation}
      \varepsilon^\text{2D}_\text{eff}(\mv{q}_\parallel) = 
       \frac{\varepsilon_1(\mv{q}_\parallel) \left[1 -\tilde{\varepsilon}_2(\mv{q}_\parallel) \tilde{\varepsilon}_3(\mv{q}_\parallel)e^{- 2 q_\parallel h} \right]}
       {1 + \left[\tilde{\varepsilon}_2(\mv{q}_\parallel) + \tilde{\varepsilon}_3(\mv{q}_\parallel)\right] e^{-q_\parallel h} + \tilde{\varepsilon}_2(\mv{q}_\parallel) \tilde{\varepsilon}_3(\mv{q}_\parallel) e^{- 2 q_\parallel h}} \label{eqn:epsmodel}. 
    \end{equation}
\end{widetext}
With this equation, the two-dimensional screened interaction (in the long-wavelength limit, see below) can be written as
    \begin{align}
      U^\text{2D}(\mv{q}_\parallel) =\frac{v^\text{2D}(\mv{q}_\parallel)}{\varepsilon^\text{2D}_\text{eff}(\mv{q}_\parallel)},
    \end{align}
where $v^\text{2D}(\mv{q}_\parallel) = v^\text{3D}(\mv{q}_\parallel, z=0)$ is the bare interaction in the 2D system.

By evaluating Eq. (\ref{eqn:epsmodel}) with the help of Eq. (\ref{eqn:eps_z_local}) we are now able to calculate the screened interaction between electrons in the free standing or embedded two dimensional monolayer \emph{directly} from the three dimensional layered bulk properties with the main approximation being the neglect of all non-localities of dielectric response in the vertical direction. We will assess the quality of this approximation in section \ref{sec:graphite_to_graphene}. 
   
  \subsection{Wannier function based formulation}
  
    The generalized Hubbard model, Eq. (\ref{eqn:Hubbard}), involves matrix elements in terms of Wannier functions.\footnote{Here, the term ``Wannier functions'' shall simply refer to localized single particle states, which form an orthormal basis of the single particle Hilbert space of interest. Possible choices include Maximally Localized Wannier Functions\cite{MLWF} or Projector Guided Wannier Functions.\cite{PWF}} Thus, the continuum electrostatic description  developed in the previous section has to be transferred to this Wannier function basis.
  
    In a situation with one Wannier orbital per unit cell and density-density type interactions only, the interaction terms of Eq. (\ref{eqn:Hubbard}) can be expressed in $k$-space, $\sum_{\mv{q}} U(\mv{q})n_{\mv{q}}n_{-\mv{q}}$, with the Fourier transformed electron density $n_{\mv{q}}=\sum_{\mv{i},\sigma} e^{i \mv{qR}_i}n_{\mv{i}\sigma}$ and the matrix elements $U(\mv{q})=\sum_{\mv{i}} e^{i \mv{qR}_i}U_{0,\mv{i}}$.\footnote{This transformation holds strictly up to an additional single particle term cancelling unphysical on-site self-interactions $-U_{00}\sum_{\mv{i},\sigma}n_{\mv{i}\sigma}$, which are implicitly contained in the k-space form of the interaction.} This kind of transformation can be made for the bare $v^\text{3D}(\mv{q})$ and the partially screened matrix elements $U(\mv{q})$. It is then natural to identify these matrix elements with the bare and screened interactions discussed in the previous section and to derive the screened Coulomb interaction of a free standing monolayer from its three 
dimensional 
layered bulk counterpart simply according to the algorithm summarized in Eqs. (\ref{eqn:epsEff}) to (\ref{eqn:epsmodel}).
  
    Often, model descriptions of real materials like graphene or transition metal dichalcogenides are more complicated as they involve multiple Wannier orbitals per unit cell and also general non-density-density interaction terms. Then we have to deal with full matrix representations of the interactions. For example, $U(\mv{q})$ is a matrix with the elements $U^{\mv{k} \mv{k'}}_{\alpha\beta\gamma\delta}(\mv{q})$, which depend in general on two initial momenta $\mv{k}$, $\mv{k'}$ and the momentum transfer $\mv{q} $ as well as four orbital indices $\alpha$, $\beta$, $\gamma$ and $\delta$. 
    
    In this case further approximations are helpful to derive from the bulk Coulomb interaction the corresponding monolayer terms. First, we neglect Coulomb assisted hopping terms between sites, which means tracing out the $\mv{k}$ and $\mv{k'}$ dependencies. Then, the Coulomb interaction depends on momentum transfer $\mv{q}$ but not on the initial momenta $\mv{k}$ and $\mv{k'}$. 
  
\begin{figure*}
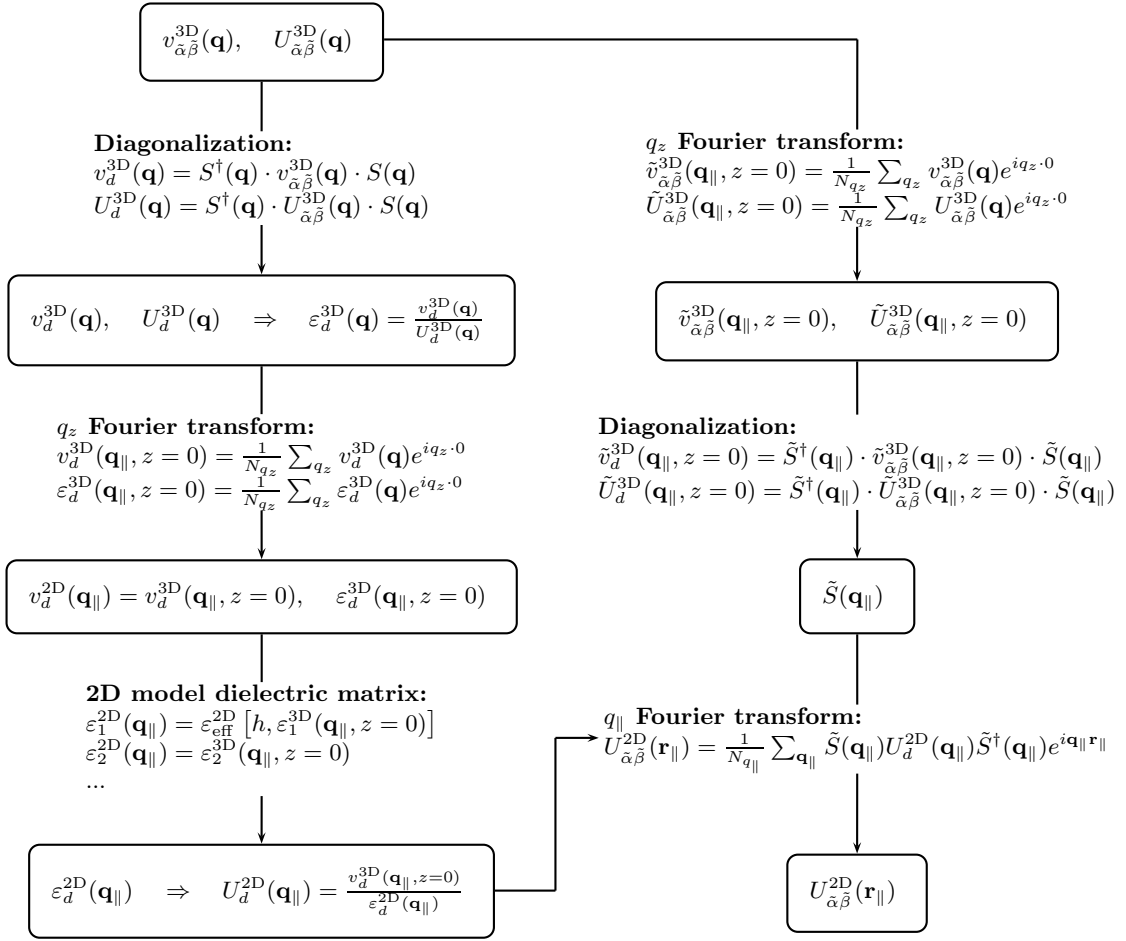

      \include{flowchart3}
      \vspace{-2cm}
      \caption{Flowchart of the Wannier function continuum electrostatics (WFCE) algorithm to obtain the screened Coulomb matrix elements of a freestanding monolayer directly from the Coulomb interaction of the corresponding layered bulk material. \label{fig:algo}}
    \end{figure*}

    To combine the macroscopic electrostatic description of the previous section with the representation in a Wannier basis, we still have to account for the orbital dependencies. We can represent the bare and screened Coulomb interaction as well as the dielectric function as quadratic matrices, using generalized indices $\tilde{\alpha} = \{\alpha, \delta\}$ and $\tilde{\beta} = \{\beta, \gamma\}$: $U_{\alpha\beta\gamma\delta}(\mv{q})\equiv U_{\tilde\alpha\tilde\beta}(\mv{q})$, which can be interpreted as interaction energies of generalized charge density waves $n_{\tilde \alpha}(\mv{q})=\sum_{\mv{k}}\langle c^\dagger_\alpha(\mv{k}+\mv{q})c_\delta(\mv{k})\rangle$ and $n_{\tilde \beta}(-\mv{q})$. The leading eigenvalue and eigenvector of $U_{\tilde\alpha\tilde\beta}(\mv{q})$ corresponds to the charge density modulations with the longest wavelength, i.e. those charge density waves where screening effects due to the environment are supposed to be strongest. By definition, continuum medium electrostatics 
describes the long wavelength response of a medium to long wavelength electric field / potential variations. We thus correct the leading eigenvalue of $U_{\tilde\alpha\tilde\beta}(\mv{q})$ according to the algorithm from the previous section, while we assume for all other eigenvalues the same screening as in the bulk. 
    The full algorithm which we refer to as ``Wannier Function Continuum Electrostatic'' (WFCE) approach can be divided into several steps, which are illustrated in the flowchart of Fig. \ref{fig:algo}. 
    
   We start with the bare, $v^\text{3D}_{\tilde{\alpha}\tilde{\beta}}(\mv{q})$, and partially screened Coulomb interactions $U^\text{3D}_{\tilde{\alpha}\tilde{\beta}}(\mv{q})$ obtained from the ab-initio calculations for the three dimensional bulk of the layered material. We then diagonalize $v^\text{3D}_{\tilde{\alpha}\tilde{\beta}}(\mv{q})$, obtain the corresponding transformation matrix $S(\mv{q})$, and assume that $S(\mv{q})$ also diagonalizes $U^\text{3D}_{\tilde{\alpha}\tilde{\beta}}(\mv{q})$. In this way, we define the 3D dielectric function
    \begin{align}
      \varepsilon_d^\text{3D}(\mv{q}) = \frac{v^\text{3D}_d(\mv{q})}{U^\text{3D}_d(\mv{q})},
      \label{eqn:app_eps_3d}
    \end{align}
  which is then by construction a diagonal matrix as indicated by the subscript index $d$ labeling the different elements on the diagonal.  We now aim to link this dielectric matrix to the interactions taking place between electrons within one monolayer and connect it to model dielectric functions derived in the context of continuum electrostatics.

  To this end, we consider the bare intralayer electronic interaction
  \begin{equation}
    v_{\tilde{\alpha} \tilde{\beta}}^{\text{3D}}(\mathbf{q_\parallel}, z=0) = \frac{1}{N_{q_z}} \sum_{q_z} v_{\tilde{\alpha} \tilde{\beta}}^{\text{3D}}(\mathbf{q_\parallel}, q_z)e^{i q_z \cdot 0}\label{eq:U2D_ab}
  \end{equation}
  and diagonalize it
  \begin{align}
    v_d^{\text{2D}}(\mathbf{q_\parallel}) = \tilde{S}^\dagger(\mathbf{q_\parallel})\cdot
	v_{\tilde{\alpha} \tilde{\beta}}^{\text{3D}}(\mathbf{q_\parallel}, z=0)\cdot
	\tilde{S}(\mathbf{q_\parallel}),
  \end{align}
  which leads to the transformation matrices $\tilde{S}(\mv{q}_\parallel)$. $N_{q_z}$ is the number of points used in the $q_z$-summation in Eq. (\ref{eq:U2D_ab}). The bare intralayer interaction is the same in the bulk layered material and in the monolayer, bilayer, etc. We assume that the transformation $\tilde{S}(\mathbf{q_\parallel})$ also diagonalizes the screened interaction within the 2D system of interest:
  \begin{align}
    U_{\tilde{\alpha} \tilde{\beta}}^{\text{2D}}(\mathbf{q_\parallel}) = \tilde{S}(\mathbf{q_\parallel}) \cdot U_d^{\text{2D}}(\mathbf{q_\parallel}) \cdot \tilde{S}^\dagger(\mathbf{q_\parallel}),
  \end{align}
  which is indeed the quanity we are searching for. What remains is to calculate 
  \begin{equation}
    U_d^{\text{2D}}(\mathbf{q_\parallel}) = \frac{v_d^{\text{2D}}(\mathbf{q_\parallel})}{\varepsilon_d^{\text{2D}}(\mathbf{q_\parallel})}.\label{eq:VD2d}
  \end{equation}
  To obtain $\varepsilon_d^{\text{2D}}$ we neglect all non-localities of the dielectric screening (c.f. Eq. \ref{eqn:app_eps_3d}) in the vertical direction, i.e. we consider $\varepsilon_d^{\text{3D}}(\mathbf{q_\parallel}, z=0) = \frac{1}{N_{q_z}} \sum_{q_z} \varepsilon_d^{\text{3D}}(\mathbf{q_\parallel}, q_z) e^{i q_z \cdot 0}$ and replace the "head element" by the model dielectric function from Eq. (\ref{eqn:epsmodel}):
  \begin{align}
    &\varepsilon_d^{\text{2D}}(\mathbf{q_\parallel}) = \label{eq:eps_matrix_2d_model} \\
    &\begin{pmatrix}
	\varepsilon_\text{eff}^\text{2D}\left[ h, \varepsilon_{1}^\text{3D}(\mathbf{q_\parallel}, z=0) \right] &         & & \\
	      & \varepsilon_{2}^{\text{3D}}(\mathbf{q_\parallel}, z=0) & & \\
	      &         & \ddots &
    \end{pmatrix}. \notag 
  \end{align}
  For simiplicity we also assume 
  \begin{equation}
    v_d^{\text{2D}}(\mathbf{q_\parallel})=v^\text{3D}_d(\mv{q}_\parallel, z=0)=\frac{1}{N_{q_z}} \sum_{q_z} v_d^{\text{3D}}(\mathbf{q_\parallel}, q_z) e^{i q_z \cdot 0}.
    \label{eq:U2d_simple}
  \end{equation}
  Since the transformation matrices $\tilde{S}(\mv{q}_\parallel)$ and $S(\mv{q})$ are not neccessarily the same, Eqs. (\ref{eq:eps_matrix_2d_model}) and (\ref{eq:U2d_simple}) imply an additional approximation, since we neglect this difference here. In future works this approximation might be removed by introducing additional transformation matrices. For the graphene systems under consideration, this approximation appears to be unproblematic as the comparison of cRPA and WFCE data in section \ref{sec:Coulomb_realsp} shows.

  From $U_{\tilde{\alpha} \tilde{\beta}}^{\text{2D}}(\mathbf{q_\parallel})$ obtained according to Eq. (\ref{eq:VD2d}) a Fourier transformation with respect to $\mv{q}_{\parallel}$ finally leads to screened Coulomb matrices element in real space 
  \begin{equation}
    U_{\tilde{\alpha}\tilde{\beta}}^{\text{2D}}(\mathbf{r_\parallel})=\frac{1}{N_{q_\parallel}} \sum_{\mathbf{q}_\parallel} U_{\tilde{\alpha} \tilde{\beta}}^{\text{2D}}(\mathbf{q_\parallel}) e^{i \mathbf{q}_\parallel \mathbf{r}_\parallel} ,
    \label{eq:}
  \end{equation}
  which can be used in extended Hubbard models like Eq. (\ref{eqn:Hubbard}). Here, $N_{q_{\parallel}}$ is the number of points used in the $q_{\parallel}$-summation.
    
\section{From Graphite to Graphene Heterostructures} \label{sec:graphite_to_graphene}

  \begin{figure*}[ht]
    \begin{center}
    
      \psfrag{screening}[c][c]{dielectric function}
    
      \psfrag{q ang}{$q$ (\AA$^{-1}$)}
      \psfrag{eps}{$\varepsilon_{1}^\text{3D}(q)$}
      \psfrag{qz = 0.1250}{$q_z = 0.125$}
      \psfrag{qz = 0.2500} {$q_z = 0.250$}
      \psfrag{qz = 0.3750}{$q_z = 0.375$}
      \psfrag{qz = 0.5000} {$q_z = 0.500$}
      \psfrag{qz = 0}{$q_z = 0$}
      \psfrag{qp = 0}{$q_\parallel = 0$}
      
      \psfrag{coulomb}[c][c]{bare and screened Coulomb interaction}
      
      \psfrag{U V (eV)}[c][c]{$v_{1}^\text{3D}(q)$, $U_{1}^\text{3D}(q)\,$(eV)}
      
      \psfrag{UU(q)abini}{$v_{\text{ab-initio}}$}
      \psfrag{VV(q)abini}{$U_{\text{ab-initio}}$}
      \psfrag{UU(q)ana}{$v_{\text{analytical}}$}
      
      \includegraphics[width=0.98\linewidth]{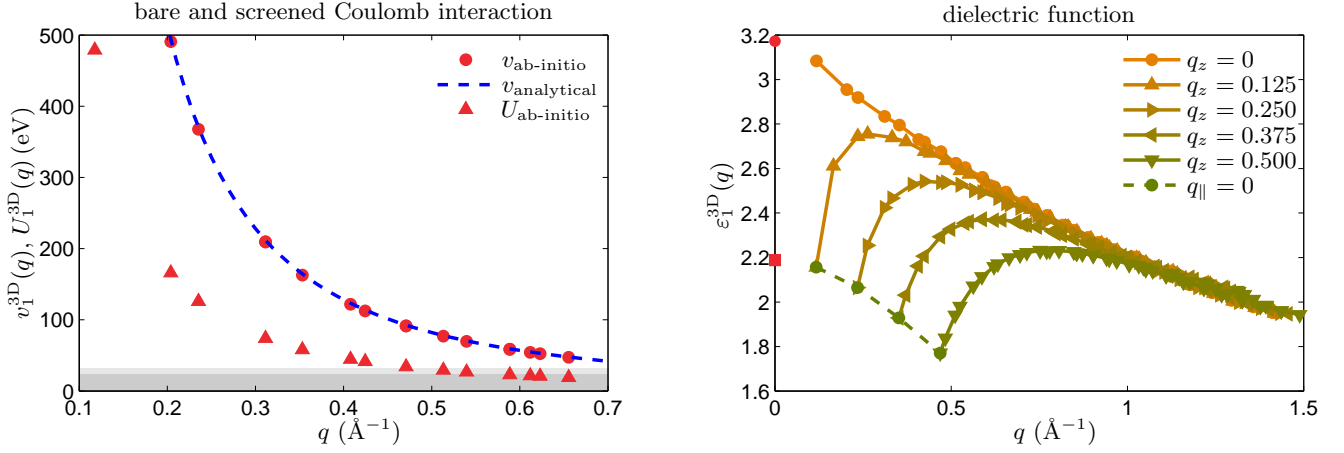}
      
      \caption{(Color online) \label{fig:ABGr}\textbf{Left:} Leading eigenvalues of the bare and screened Coulomb matrix elements of graphite for $q_z = 0$ obtained from ab-initio calculations together with the analytical description of the unscreened interaction (dashed blue line). The (light) gray area indicate the interval of all other eigenvalues of the (bare) screened Coulomb matrix.  \textbf{Right:} Momentum dependent leading eigenvalue of dielectric function of graphite obtained from cRPA calculations for different values of $q_z$. The red markers for $q = 0$ indicate the parallel (circle) and perpendicular (square) limits of the screening. }
      
    \end{center}
  \end{figure*}

  We derive effective Coulomb interactions of monolayer graphene (MLG), bilayer graphene (BLG) and  Ir intercalated graphite (Gr/Ir) from the interaction calculated for bulk graphite in the WFCE approach and benchmark our results against direct ab-initio calculations for these systems as well as analytical expressions which are valid in the long wavelength limit. In all cases, we consider Coulomb matrix elements in terms of Wannier functions for the carbon $p_z$ orbitals.
  
  \subsection{Bulk Graphite}
      
    The left panel of Fig. \ref{fig:ABGr} shows the leading eigenvalue of the bare and screened Coulomb interaction in AB stacked graphite, which plays the role of the initial bulk material, here. The former is perfectly interpolated by the analytic expression
    \begin{align}
      v_{1}(\mv{q}) = \frac{4 \pi e^2}{\tilde{V}} \frac{1}{q^2}
    \end{align}
    which is illustrated in Fig. \ref{fig:ABGr} for the $q_z = 0$ direction. Here, $e$ is the elementary charge and $\tilde{V}$ is the volume per atom. The fact that the leading eigenvalue of the bare Coulomb interaction in terms of Wannier orbitals matches the long wavelength continuum description within large parts of the Brillouin zone very closely motivates us to consider exactly this part of the Coulomb interaction in the WFCE approach.

    The leading bare interaction eigenvalue is basically independent of any microscopic properties. This is different in the case of the screened interaction. Here, microscopic and macroscopic properties are involved through the dielectric screening of the real material background, as can be seen from the right panel of Fig. \ref{fig:ABGr}. In the limit of small $q=|\mv{q}|\to 0$ the tensorial character of the dielectric function becomes obvious. Here, we find $\varepsilon_{\parallel}\approx 3.2$ for the in-plane fields and $\varepsilon_{\perp}\approx 2.2$ for the out-of-plane direction. At larger momentum transfer $q$, the direction dependence of the dielectric function is less pronounced. Besides the leading eigenvalues of the Coulomb matrices the energetic interval of the other eigenvalues are marked by the (light) gray shaded areas in the left panel of Fig. \ref{fig:ABGr} for the (bare) screened interaction. These matrix elements correspond to electronic density variations within the unit cell and 
correspondingly short wavelengths.

  \subsection{Freestanding mono- and bilayer graphene}

    We derive the screened Coulomb interactions in freestanding mono- and bilayer graphene using the bulk graphite data in the WFCE approach. The leading eigenvalues of the Coulomb interaction and the corresponding effective dielectric functions are shown as function of momentum transfer in Fig. \ref{fig:Graphene}. While the comparison of results from direct ab-initio calculations and from the WFCE approach reveals generally very good agreement, there are some systematic deviations between both approaches in the limit of $q\to 0$. To understand the origin of these deviations it is instructive to compare the bare Coulomb interaction matrix elements obtained from WFCE and ab-intio results to the analytical expression for the bare Coulomb interaction between electrons confined to a two-dimensional film in the long-wavelength limit:

    \begin{figure*}[ht]
      \begin{center}
	\psfrag{q ang}{$q$ (\AA$^{-1}$)}
	\psfrag{eps}{$\varepsilon_{1}^\text{2D}(\mv{q}_\parallel)$}
	
	\psfrag{UV(eV)}[c][c][0.7]{$v_{1}^\text{2D}(q)$, $U_{1}^\text{2D}(q)\,$(eV)}
	
	\psfrag{monolayer}[c][c]{monolayer graphene}
	\psfrag{bilayer}[c][c]{bilayer graphene}
	
	\psfrag{UU(q)abini}{$v_{\text{ab-initio}}^\text{2D}$}
	\psfrag{VV(q)abini}{$U_{\text{ab-initio}}^\text{2D}$}
	\psfrag{eps(q)abini}{$\varepsilon_{\text{ab-initio}}^\text{2D}$}
	
	\psfrag{UU(q)WFCE}{$v_{\text{WFCE}}^\text{2D}$}
	\psfrag{VV(q)WFCE}{$U_{\text{WFCE}}^\text{2D}$}
	\psfrag{eps(q)WFCE}{$\varepsilon_{\text{WFCE}}^\text{2D}$}
	
	\psfrag{UU(q)ana}{$v_{\text{analytical}}^\text{2D}$}
	
	\includegraphics[width=0.98\linewidth]{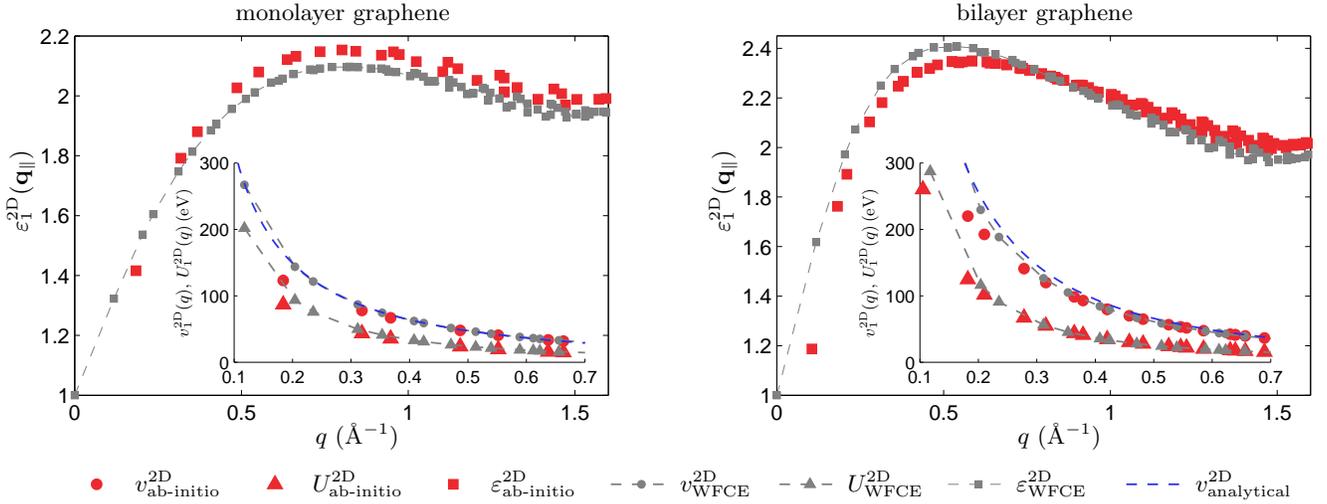}
	\caption{(Color online) Leading eigenvalue of the dielectric function (outer frame), bare and screened Coulomb matrices (inner frame) of monolayer (\textbf{left}) and bilayer graphene (\textbf{right}). Red markers indicate ab-initio calculations and gray markers the WFCE values. \label{fig:Graphene}}
      \end{center}
    \end{figure*}

    \begin{align}
      v_{1}^\text{2D}(\mv{q}_\parallel) &= \frac{h}{2 \pi} \int_{-\pi/h}^{+\pi/h}  \frac{4 \pi e^2}{\tilde{V}} \frac{1}{q^2} dq_z \label{eqn:Bare2DU} \\
      &= \frac{4 e^2}{\tilde{A}} \frac{\arctan\left( \frac{\pi}{q_\parallel h} \right) }{q_\parallel}\notag
    \end{align}
    where $\tilde{A}$ is the unit cell area per atom and the effective height $h$ can be chosen to be the interlayer distance $h=d \approx 3.35\,$\AA\ for the monolayer and $h=2d$ for the bilayer. At small momentum transfer ($q_\parallel h \ll 1$), this bare interaction approaches the well known limit $v_{1}^\text{2D}(\mv{q}_\parallel)\to\frac{2\pi e^2}{\tilde A q}$. The term $\arctan\left( \frac{\pi}{q_\parallel h} \right)$
    in Eq. (\ref{eqn:Bare2DU}) plays the role of a ``form factor'' which accounts for the effective height $h$ of the two dimensional layer. For, both, the monolayer and the bilayer the WFCE results match the analytic expression, which becomes exact for $q\to 0$, almost perfectly, in contrast to the ab-initio data. The ab-initio calculations performed here are in fact supercell calculations with periodic boundary conditions. This creates periodic images of the mono- and bilayer graphene in $z$ direction, which contribute to the effective screening. We employ an extrapolation to infinite supercell height (see Appendix) in order to obtain the monolayer limit, which becomes somewhat inaccurate for small q. Thus, the deviation of ab-initio and WFCE Coulomb matrix elements as well as dielectric functions at small $q$ is likely due to this extrapolation problem in the ab-initio data.
  
    At intermediate $q$ the WFCE and the ab-intio dielectric function are in very good agreement. For both, mono- and bilayer graphene, the screening rises from $1$ to a maximum at intermediate $q$ and slightly decreases afterwards towards the edges of the first Brillouin zone \cite{Thygesen_PRB2013}. Here, the non-locality ($q$-dependence) of the screening becomes clearly visible. In the long wavelength limit the screening vanishes, since we are dealing with a free standing two dimensional layer, which is embedded in an infinite three dimensional vacuum. By decreasing the wavelength, or increasing $\mv{q}_\parallel$, the Coulomb interaction starts to be screened like in a three dimensional bulk system, which manifests as an increased value of the dielectric function. The main differences between the effective dielectric functions in mono- and bilayer graphene is the gradient towards the intermediate maximum and the absolute value of the maximum, which are steeper and higher, respectively, in the bilayer. I. 
e. the 
long range Coulomb 
interaction is less screened in the monolayer than in the bilayer, while the short range screening is more or less the same. The screened Coulomb interaction obtained from WFCE interpolates the corresponding cRPA data very well, as can be seen from Fig. \ref{fig:Graphene}. Thus, we have proven, that the WFCE approach to calculate the two dimensional Coulomb repulsion directly from the three dimensional bulk data without introducing additional parameters works very well. 
  
  \subsection{Graphene in a metallic surrounding\label{sec:GrIr}}

    \begin{figure}[ht]
      \begin{center}
      
	\psfrag{q ang}{$q$ (\AA$^{-1}$)}
	\psfrag{eps}{$\varepsilon_{1}^\text{2D}(q)$}
	\psfrag{U V (eV)}[c][c][0.7]{$v_{1}^\text{2D}(q)$, $U_{1}^\text{2D}(q)\,$(eV)}
	
	\psfrag{UU(q)abini}{$v_{\text{ab-initio}}$}
	\psfrag{VV(q)abini}{$U_{\text{ab-initio}}$}
	\psfrag{EE(q)abini}{$\varepsilon_{\text{ab-initio}}$}
	
	\includegraphics[width=0.96\linewidth]{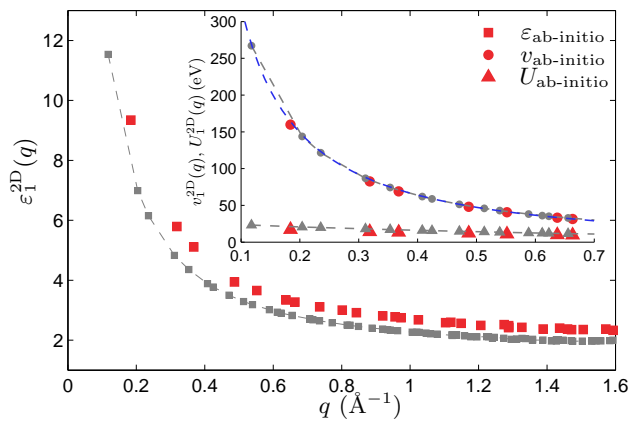}
	\caption{(Color online) Leading eigenvalue of the dielectric function (outer frame), bare and screened Coulomb matrices (inner frame) of Gr/Ir. Red markers indicate cRPA calculations and gray markers the values from the WFCE approach. \label{fig:GrIr}}
      \end{center}
    \end{figure}
    
    Regarding the change in electronic interactions, the opposite extreme case to going from bulk graphite to free standing monolayer is the case of graphene embedded in some metallic environment. Perfect metallic screening by the environment corresponds to $\varepsilon_2, \varepsilon_3 \to \infty$, in contrast to the case of $\varepsilon_2= \varepsilon_3=1$ for monolayers surrounded by vacuum. In experiment, graphene is frequently grown on metals like Ir\cite{ndiaye_two-dimensional_2006} or Cu\cite{gao_epitaxial_2010} or can be surrounded by metals e.g. in graphite intercalation compounds\cite{dresselhaus_intercalation_2002}. We consider Coulomb interactions in graphene surrounded by Ir in the following.

    To this end, we calculated Coulomb interactions for a periodically repeated slab composed of graphene monolayer and one ``monolayer'' of iridium by means of cRPA. This system can also be interpreted as Ir intercalated graphite. To model this system with the WFCE approach we assume perfect metallic screening by Ir, $\varepsilon_2, \varepsilon_3 \to \infty$, and use the effective height {$h=3.35$~\AA} of graphene, as before. The resulting leading eigenvalues of the Coulomb interaction within the carbon $p_z$ Wannier orbitals as well as the corresponding effective dielectric function  are shown in Figure \ref{fig:GrIr}. The metallic surrounding leads to diverging $\varepsilon^{2D}_{1}(\mv{q})$ at long wavelengths $\mv{q} \rightarrow 0$ in the cRPA and in the WFCE approach, as it must be. In contrast to free standing mono- and bilayer graphene the screened interactions in Gr/Ir do not diverge at small $q$, where the Coulomb interaction is now efficiently screened by the metallic environment. 
  
    The overall characteristics of interactions and screening as obtained from WFCE agree with the cRPA calculations. Nevertheless there is a systematic underestimation of the screening $\varepsilon^{2D}_{1}(\mv{q})$ on the order of $\approx17\%$ by the WFCE approach as compared to the cRPA. Hence, the screened Coulomb interactions are correspondingly overestimated by WFCE, here. On physical grounds it is clear that the WFCE approach can become inaccurate when there is hybridization between e.g. a monolayer of graphene and some metallic surrounding. In this case the assignment of an effective height $h$ to the graphene layer and a separation into a subsystem of graphene and "the environment" is ambiguous. The underestimation of the screening in the WFCE approach can indeed be cured by decreasing the effective height to $h\approx2.8\,$\AA\ of the modeled monolayer. 
    Treating $h$ as an adjustable parameter, that is derived from e.g. cRPA calculations is one possibility if for instance very complex heterostructures shall be considered and the intercalated system is used as the bulk starting point in WFCE. Here, we are taking graphite as the bulk starting point to treat Ir intercalated graphite and keep $h=3.35$\AA\ to stay with a parameter free model.
  
  \subsection{Coulomb Interactions in Real Space}\label{sec:Coulomb_realsp}
  
\begin{figure}[ht]
      \begin{center}
      
	\psfrag{r ang}{$r$ (\AA)}
	\psfrag{V (eV)}[c][c]{$U^\text{2D}(r)$ (eV)}
	
	\psfrag{graphite}{graphite}
	\psfrag{monolayer}{MLG}
	\psfrag{bilayer}{BLG}
	\psfrag{GrIr}{Gr/Ir}
	
	\includegraphics[width=0.96\linewidth]{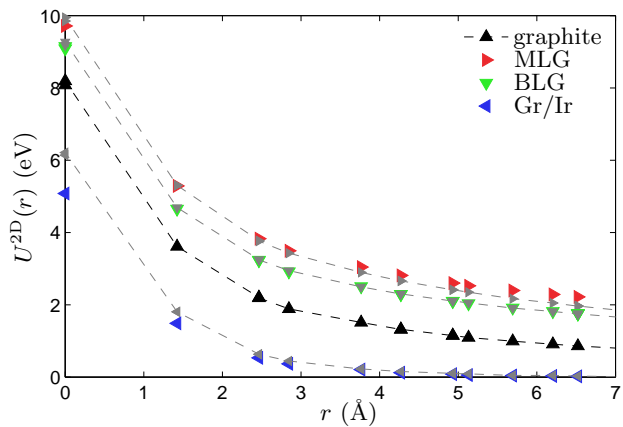}
	\caption{(Color online) Density-density matrix elements of the screened Coulomb interactions for graphite, monolayer and bilayer graphene as well as Gr/Ir in real space. Colored markers show ab-initio values, gray markers connected by dashed lines show the WFCE values. \label{fig:VVReal}}
	
      \end{center}
    \end{figure}

    In order to use the Coulomb terms obtained within the WFCE approach in a generalized Hubbard model, in which interactions matrix elements enter in real space representation, we perform a Fourier transformation:
      \begin{align}
	U_{\tilde{\alpha} \tilde{\beta}}^{\text{2D}}(\mathbf{r_\parallel}) = \frac{1}{N_{q_{\parallel}} } \sum_{\mv{q}_\parallel} U_{\tilde{\alpha} \tilde{\beta}}^{\text{2D}}(\mathbf{q_\parallel}) e^{i \mv{q}_\parallel \cdot \mv{r}_\parallel}
      \end{align}
    In the case of graphite an additional sum over the $q_z$ component is performed. The resulting values for density-density like $U_{\tilde{\alpha} \tilde{\beta}}(\mathbf{r_\parallel})$ as obtained from cRPA and WFCE are given in Table \ref{tab:bilayer} and depicted in Fig. \ref{fig:VVReal} for mono- and bilayer graphene as well as graphite and the Gr/Ir system. 
  
    \begin{table}[ht]
      \caption{\label{tab:bilayer}cRPA and WFCE screened Coulomb interactions for graphite, monolayer and bilayer graphene and Ir intercalated graphite in eV. Since the AB stacking breaks the sublattice symmetry in bilayer graphene for every second neighbor, some interactions are given separately for the A and B sublattice (in one of the two layers).}
      \begin{ruledtabular}
	\begin{tabular}{lllllll}
	  System 	& $U_0$ & $U_1$ & $U_2$ & $U_3$ & $U_4$ & $U_5$ \\ \hline
	  Graphite 	& $8.1 / 8.2$ & $3.6$ & $2.2 / 2.2$ & $1.9$ & $1.5 / 1.5$ & $1.3$ \\ \hline
	  MLG \\
	  cRPA		& $9.7$ & $5.3$ & $3.8$ & $3.5$ & $3.0$ & $2.8$ \\
	  WFCE 		& $9.9$ & $5.3$ & $3.8$ & $3.4$ & $2.9$ & $2.6$ \\ \hline
	  BLG\\
	  cRPA 		& $9.1 / 9.2$ & $4.7$ & $3.2 / 3.2$ & $2.9$ & $2.5 / 2.5$ & $2.3$ \\
	  WCFE 		& $9.2 / 9.3$ & $4.7$ & $3.3 / 3.2$ & $3.0$ & $2.5 / 2.5$ & $2.3$ \\ \hline
	  Gr/Ir \\
	  cRPA 		& $5.1$ & $1.5$ & $0.5$ & $0.4$ & $0.2$ & $0.1$ \\
	  WCFE 		& $6.2$ & $1.8$ & $0.6$ & $0.4$ & $0.2$ & $0.2$ \\
	\end{tabular}
      \end{ruledtabular}
    \end{table}

The screened Coulomb interaction in monolayer graphene is over the whole $r_\parallel$ range bigger than the corresponding values of bilayer graphene, graphite, and Ir intercalated graphene. Since the bare Coulomb interactions (not shown here) are nearly the same in all cases, variations of the background screened interactions are almost entirely due to the successively stronger screening when going from monolayer graphene via bilayer graphene and graphite to graphene encapsulated in a metal.

    In agreement with Ref. \onlinecite{wehling_strength_2011}, we find sizeable nonlocal effective Coulomb interactions for graphene, bilayer graphene and graphite, which can be however strongly reduced due to screening by the environment. This can be seen from comparison to the Gr/Ir case. Here, the Coulomb interaction is strongly reduced at all $r_\parallel$ under consideration, i.e. by about a factor of 2 for the local terms and more than a factor of 10 for interaction terms beyond fourth nearest neighbours. 
  
    The comparison of effective Coulomb interaction obtained from direct cRPA and WFCE calculations shows generally very good quanitative agreement with deviations of less than $10\%$. The only exception are in the Gr/Ir data. Here, the local Coulomb interactions are overestimated by WFCE by about $1$\,eV, which is likely a result of the approximated effective monolayer height, as dicussed in section \ref{sec:GrIr}. Nevertheless, even in this "worst-case" the WFCE approach accounts for $\sim 80\%$ of the increased screening provided by the metallic environment. 
  
  \section{Electronic Ground State of Bilayer Graphene Heterostructures}\label{sec:bilayer}

    \begin{figure*}[ht]
      \begin{center}
      
	\psfrag{coulomb}[c][c]{screened Coulomb interaction}
	\psfrag{screening}[c][c]{dielectric function}
      
	\psfrag{r ang}{$r$ (\AA)}
	\psfrag{q ang}{$q$ (\AA$^{-1}$)}
	\psfrag{V (eV)}[c][c]{$U^\text{2D}(r)$ (eV)}
	\psfrag{eps}{$\varepsilon_{1}^\text{2D}(q)$}
	
	\psfrag{Eps 1 1}{($1$ / $1$)}
	\psfrag{Eps 1 5}{($1$ / $5$)}
	\psfrag{Eps 1 inf}{($1$ / $\infty$)}
	\psfrag{Eps inf inf}{($\infty$ / $\infty$)}
	
	\includegraphics[width=0.98\linewidth]{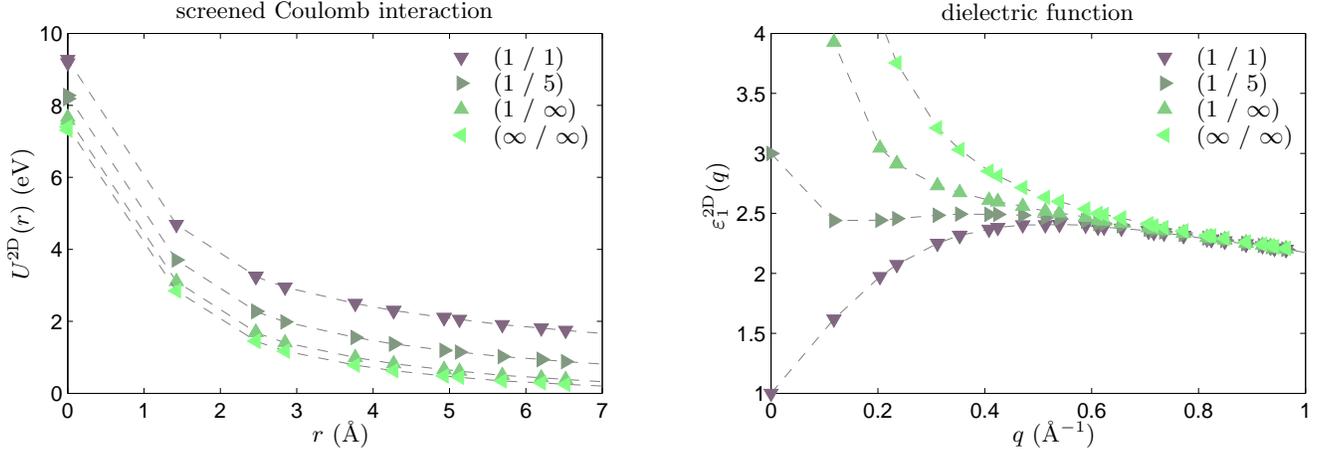}

	\caption{(Color online) \label{fig:Bilayer}\textbf{Left:} WFCE density-density matrix elements of the background screened Coulomb interactions for bilayer graphene in real space for different dielectric surroundings ($\varepsilon_2$ / $\varepsilon_3$). \textbf{Right:} Corresponding dielectric functions in momentum space. }
      \end{center}
    \end{figure*}
    
    Bilayer graphene is known to host competing symmetry broken electronic ground states. Theoretical studies have predicted that charge- and spin-density waves (CDW or SDW), quantum spin hall states (QSH), nematic, superconducting and excitonic insulator states could emerge in the bilayer \cite{song_excitonic_2012, scherer_instabilities_2012, nilsson_electron-electron_2006, nandkishore_dynamical_2010, lemonik_competing_2012, lang_antiferromagnetism_2012, kotov_electron-electron_2012}.  
    Different experiments have addressed the issue of symmetry broken ground states in bilayer graphene \cite{weitz_broken-symmetry_2010, mayorov_interaction-driven_2011, martin_local_2010, jr_transport_2012, freitag_spontaneously_2012, feldman_broken-symmetry_2009} but the issue remains controversial also from the experimental point of view and it is e.g. unclear whether or not the ground state exhibits a finite electronic excitation gap. In the end, it appears very likely that microscopic material specific details of the effective interactions determine which electronic phases are realized.   

    As a dielectric substrate or also some metallic environment can provide additional screening of the effective Coulomb interactions in bilayer, we investigate how different types of environments affect the electronic ground state of bilayer graphene. To this end we use the WFCE approach to study the influence of a dielectric substrate ($\varepsilon_3 = 5$; $\varepsilon_2 = 1$) and a metallic substrate ($\varepsilon_3 \to \infty$; $\varepsilon_2 = 1$) as well as a metallic encapsulation ($\varepsilon_2 =\varepsilon_3 \to\infty$). The results can be seen in Fig. \ref{fig:Bilayer} as well as in Tab. \ref{tab:bilayer_m}. 

    The effective dielectric function diverges for $q \rightarrow 0$ in the case of the metallic substrate/environment, whereas a finite $\varepsilon_{1}^\text{2D}(q=0) > 1$ can be found for the dielectric substrate, as it must be. The resulting effective Coulomb interactions can be clearly reduced due to environmental screening as comparison with the free standing bilayer data demonstrates. 
    To understand the resulting effects on the electronic ground states, we make use of an electronic phase diagram that is based on a recent functional renormalization group study \cite{scherer_instabilities_2012} and that gives the ground states in dependence of the local, nearest-, and next-nearest neighbor screened Coulomb interaction. 
    The interaction strengths obtained here put the ground state of bilayer graphene in all environments considered on the Antiferromagnetic (AF)-SDW side of a phase transition line between the QSH and the AF-SDW phases found in Ref. \onlinecite{scherer_instabilities_2012}. 
    This result of the AF-SDW being stable in all situations is quite surprising given the fact, that the Coulomb interactions are changed quite drastically.
  
    \begin{table}[ht]
      \caption{\label{tab:bilayer_m}Screened Coulomb interaction for bilayer graphene modified through different dielectric environments $\varepsilon_{2,3}$. All values are given in units of eV. Since the AB stacking breaks the sublattice symmetry for every second neighbor, some interactions are given separately for the A and B sublattice (in one of the two layers).}
      \begin{ruledtabular}
	\begin{tabular}{ccllll}
	$\varepsilon_2$ 	& $\varepsilon_3$& $U_0$ & $U_1$ & $U_2$ & $U_3$ \\ \hline
	$1$ & $1$ 		& $9.2 / 9.3$	& $4.7$	& $3.3 / 3.2$	& $3.0$	 \\
	$1$ & $5$ 		& $8.2 / 8.3$	& $3.7$	& $2.3 / 2.3$	& $2.0$	 \\
	$1$ & $\infty$		& $7.6 / 7.7$	& $3.1$	& $1.7 / 1.7$	& $1.4$	 \\
	$\infty$ & $\infty$ 	& $7.3 / 7.4$	& $2.9$	& $1.5 / 1.5$	& $1.2$	 \\
	\end{tabular}
      \end{ruledtabular}
    \end{table}
  
\section{Conclusion}

  We have established a scheme that combines cRPA calculations of layered materials in the bulk and continuum medium electrostatics to derive effective Coulomb interaction matrix elements in terms of Wannier functions for free standing 2D materials as well as 2D materials embedded in complex dielectric environments. We call the scheme "Wannier function continuum electrostatics" (WFCE) approach. It allows us to avoid supercell calculations involving complex environments or large vacuum volumes on the ab-initio side, which are numerically very costly in implementations using periodic boundary conditions. While the WFCE approach might be further improved and generalized in the future, already the simplest version presented here predicts effective Coulomb matrix elements for monolayer and bilayer graphene very accurately, i.e. for instance the local Hubbard interaction agrees with the full cRPA calculation within $0.2$~eV. 

  Our modelling shows that Coulomb interactions can be strongly manipulated in 2D materials like graphene by means of screening provided by the environments which can be substrates, adsorbates or other 2D materials. Given the numerical simplicity of the WFCE approach, we anticipate that it could be very useful in the context of materials design, as effects of different kinds of dielectric environments on Coulomb interactions in layered materials can be modelled quickly and quantitatively, now.

  \acknowledgments
The authors thank M. Katsnelson and A. Lichtenstein for discussions and acknowledge financial support from the DFG via FOR 1346 and SPP 1459 as well as the European Graphene Flagship.

\appendix

\section{Computational Details of the Ab-initio Simulations\label{sec:appendix_abinitio}}
  
\begin{figure}[ht]
    \begin{center}
      \subfloat[][]{\includegraphics[width=0.45\linewidth]{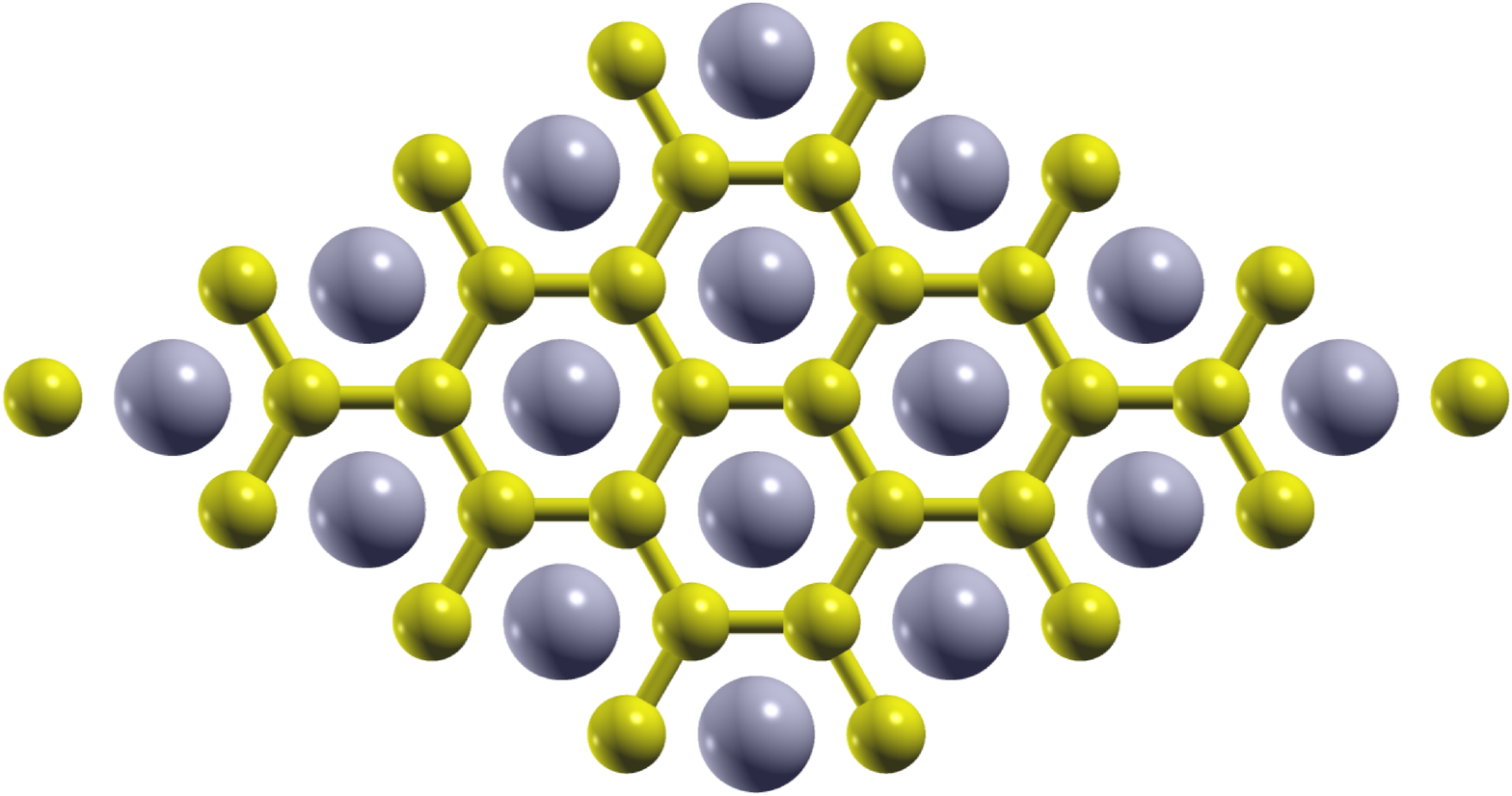}}
      \qquad
      \subfloat[][]{\includegraphics[width=0.33\linewidth]{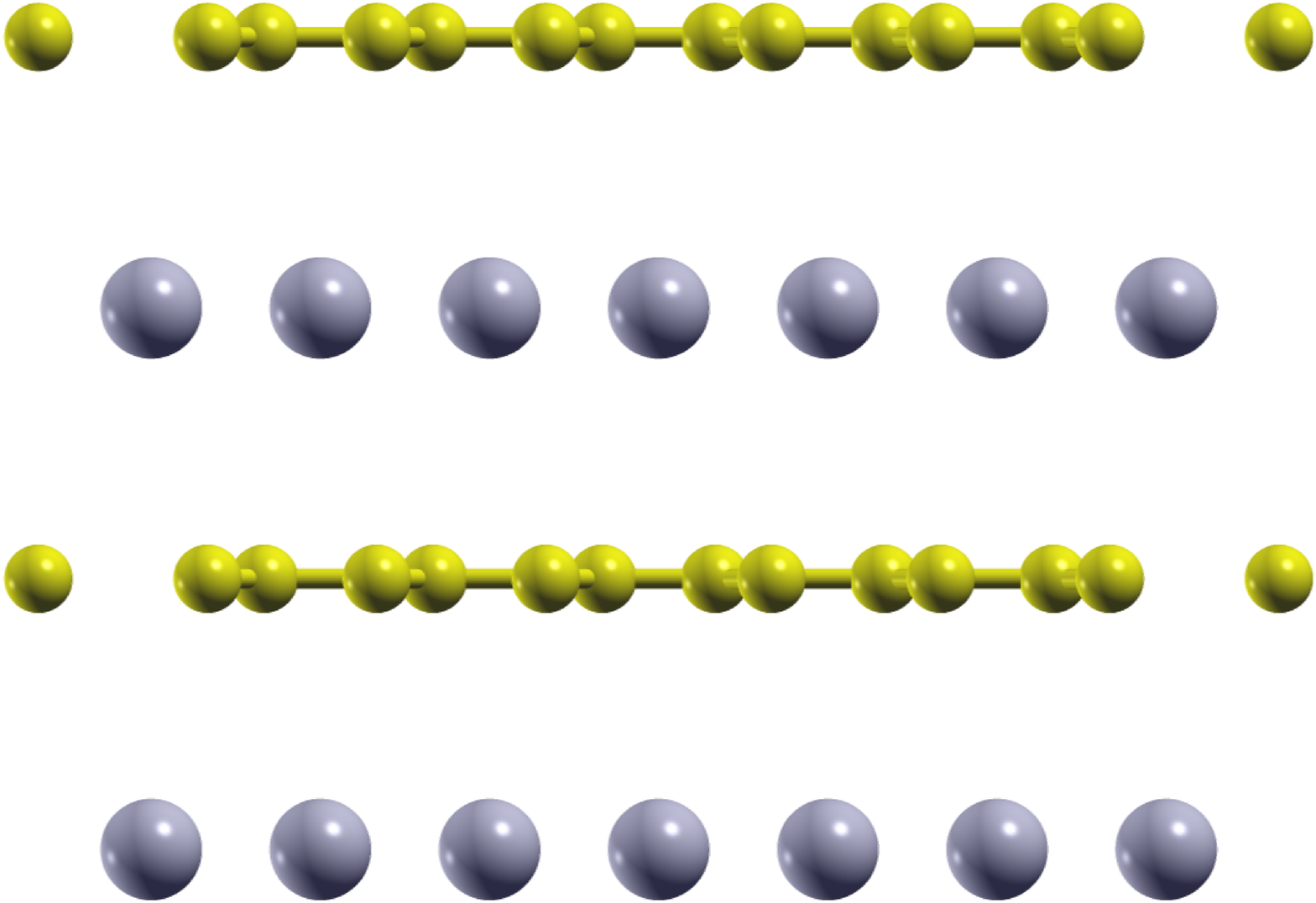}}
      \caption{\label{fig:grir}\textbf{(a)} Top and \textbf{(b)} side view of the Gr/Ir system.}
    \end{center}
  \end{figure}

  In all calculations the carbon nearest neighbor distance is set to $1.42\,$\AA. In graphite and BLG an AB Bernal stacking is chosen with an interlayer distance of $3.35\,$\AA \cite{baskin_lattice_1955}.  To model Ir intercalated graphite we use the structure depicted in Fig. \ref{fig:grir}. The distance between the graphene layers and the Ir layer is set to the experimental value of $3.41\,$\AA \cite{busse_graphene_2011}.
  
  For all DFT calculations the generalized gradient approximation (PBE) \cite{perdew_generalized_1996} is used. In the case of carbon atoms we use an angular momentum cut-off of $l_\text{cut} = 6$ and $l_\text{cut} = 8$ for iridium. The plane-wave cut-off is set to $4.5\,a_0^{-1}$, where $a_0$ is the Bohr radius. The involved $k$ meshes and the energy cut-offs for the polarization function are shown in Tab. \ref{tab:systemdetails}. The energy cut-off  corresponds to the energy of the highest, unoccupied band (and thus to the total number of empty bands) involved in the calculation of the polarization function. Since in the case of MLG and BLG several ``vacuum distances'' (see below) have been used, the number of empty bands had to be adjusted for each vacuum height (corresponding to the given energy cut-off).
  
  A ``vacuum distance'' $h_\text{vac}$ (the distance between adjacent layers) is introduced, since we embed the mono- or bilayer in a three dimensional unit cell. Thereby, we produce, due to the periodic boundaries, an infinite stack of mono- or bilayers separated by the unit cell height.
  The freestanding situation is obtained in the limit of $h_\text{vac} \rightarrow \infty$. To approximate this limit, we do several calculations for different vacuum distances (ranging from $h_\text{vac} \approx 15\,$\AA\ to $h_\text{vac} \approx 30\,$\AA) and extrapolate the freestanding value $U_{\alpha\beta}(\mv{q}, \infty)$ by fitting the results to
  \begin{align}
    U_{\alpha\beta}(\mv{q}, h_\text{vac}) = U_{\alpha\beta}(\mv{q}, \infty) + \frac{b_{\alpha\beta}(\mv{q})}{h_\text{vac}}.
  \end{align}
  
  \begin{table}[ht]
    \caption{\label{tab:systemdetails}Ab-initio details for each system. The polarization energy cut-offs are given relative to the graphene Dirac-cone position.}
    \begin{ruledtabular}
      \begin{tabular}{cccl}
	system 		& DFT $k$ mesh 			& cRPA $k$ mesh 			& energy cut-off  \\ \hline
	AB graphite	& $16 \times 16 \times 5$	& $25 \times 25 \times 8$	& $\approx 60\,$eV \\
	MLG		& $14 \times 14 \times 1$	& $16 \times 16 \times 1$	& $\approx 120\,$eV \\
	BLG		& $14 \times 14 \times 1$	& $28 \times 28 \times 1$	& $\approx 60\,$eV \\
	Gr/Ir		& $16 \times 16 \times 5$	& $28 \times 28 \times 1$	& $\approx 180\,$eV \\
      \end{tabular}
    \end{ruledtabular}
  \end{table}

\bibliography{references}

\end{document}

%% file: flowchart3.tex
    \begin{psmatrix}[rowsep=0.6,colsep=0.5]
    
      \psframebox[linearc=0.14,cornersize=absolute,framesep=8pt]{\small
	$v_{\tilde{\alpha} \tilde{\beta}}^{\text{3D}}(\mathbf{q})$, 
	$\quad U_{\tilde{\alpha} \tilde{\beta}}^{\text{3D}}(\mathbf{q})$
      }
      
      &
      &\\
    
      \tabular{l}\small
	\textbf{Diagonalization:}\\
	$v_d^{\text{3D}}(\mathbf{q}) = S^\dagger(\mathbf{q}) \cdot v_{\tilde{\alpha} \tilde{\beta}}^{\text{3D}}(\mathbf{q}) \cdot S(\mathbf{q})$\\
	$U_d^{\text{3D}}(\mathbf{q}) = S^\dagger(\mathbf{q}) \cdot U_{\tilde{\alpha} \tilde{\beta}}^{\text{3D}}(\mathbf{q}) \cdot S(\mathbf{q})$
      \endtabular
      
      &
      &
      
      \tabular{l}\small
	\textbf{$q_z$ Fourier transform:}\\
	$\tilde{v}_{\tilde{\alpha} \tilde{\beta}}^{\text{3D}}(\mathbf{q_\parallel}, z=0) = \frac{1}{N_{q_z}} \sum_{q_z} v_{\tilde{\alpha} \tilde{\beta}}^{\text{3D}}(\mathbf{q})e^{i q_z \cdot 0}$\\
	$\tilde{U}_{\tilde{\alpha} \tilde{\beta}}^{\text{3D}}(\mathbf{q_\parallel}, z=0) = \frac{1}{N_{q_z}} \sum_{q_z} U_{\tilde{\alpha} \tilde{\beta}}^{\text{3D}}(\mathbf{q})e^{i q_z \cdot 0}$\\
      \endtabular\\
      
      \psframebox[linearc=0.14,cornersize=absolute,framesep=8pt]{\small
	$v_d^{\text{3D}}(\mathbf{q})$, 
	$\quad U_d^{\text{3D}}(\mathbf{q})$
	$\quad \Rightarrow \quad \varepsilon_d^{\text{3D}}(\mathbf{q}) = \frac{v_d^{\text{3D}}(\mathbf{q})}{U_d^{\text{3D}}(\mathbf{q})}$
      }
      
      &
      &
      
      \psframebox[linearc=0.14,cornersize=absolute,framesep=8pt]{\small
	$\tilde{v}_{\tilde{\alpha} \tilde{\beta}}^{\text{3D}}(\mathbf{q_\parallel}, z=0)$,
	$\quad \tilde{U}_{\tilde{\alpha} \tilde{\beta}}^{\text{3D}}(\mathbf{q_\parallel}, z=0)$
      }\\
      
      \tabular{l}\small
	\textbf{$q_z$ Fourier transform:}\\
	$v_d^{\text{3D}}(\mathbf{q_\parallel}, z=0) = \frac{1}{N_{q_z}} \sum_{q_z} v_d^{\text{3D}}(\mathbf{q}) e^{i q_z \cdot 0}$\\
	$\varepsilon_d^{\text{3D}}(\mathbf{q_\parallel}, z=0) = \frac{1}{N_{q_z}} \sum_{q_z} \varepsilon_d^{\text{3D}}(\mathbf{q}) e^{i q_z \cdot 0}$\\
      \endtabular
      
      &
      &
      
      \tabular{l}\small
	\textbf{Diagonalization:}\\
	$\tilde{v}_d^{\text{3D}}(\mathbf{q_\parallel}, z=0) = \tilde{S}^\dagger(\mathbf{q_\parallel}) \cdot \tilde{v}_{\tilde{\alpha} \tilde{\beta}}^{\text{3D}}(\mathbf{q_\parallel}, z=0) \cdot \tilde{S}
	(\mathbf{q_\parallel})$\\
	$\tilde{U}_d^{\text{3D}}(\mathbf{q_\parallel}, z=0) = \tilde{S}^\dagger(\mathbf{q_\parallel}) \cdot \tilde{U}_{\tilde{\alpha} \tilde{\beta}}^{\text{3D}}(\mathbf{q_\parallel}, z=0) \cdot \tilde{S}
	(\mathbf{q_\parallel})$
      \endtabular\\
      
      \psframebox[linearc=0.14,cornersize=absolute,framesep=8pt]{\small
	$v_d^{\text{2D}}(\mathbf{q_\parallel}) = v_d^{\text{3D}}(\mathbf{q_\parallel}, z=0)$, 
	$\quad \varepsilon_d^{\text{3D}}(\mathbf{q_\parallel}, z=0)$
      }
      
      &
      &
      
      \psframebox[linearc=0.14,cornersize=absolute,framesep=8pt]{\small
	$\tilde{S}(\mathbf{q_\parallel})$
      }\\
      
      \tabular{l}\small
	\textbf{2D model dielectric matrix:}\\
	$\varepsilon_{1}^{\text{2D}}(\mathbf{q_\parallel}) = \varepsilon_\text{eff}^{\text{2D}}\left[ h, \varepsilon_{1}^{\text{3D}}(\mathbf{q_\parallel}, z=0) \right]$\\
	$\varepsilon_{2}^{\text{2D}}(\mathbf{q_\parallel}) =  \varepsilon_{2}^{\text{3D}}(\mathbf{q_\parallel}, z=0)$\\
	$...$\\
      \endtabular
      
      &
      
      &
      \tabular{l}\small
	\textbf{$q_\parallel$ Fourier transform:}\\
	$U_{\tilde{\alpha} \tilde{\beta}}^\text{2D}(\mathbf{r}_\parallel) = \frac{1}{N_{q_\parallel}} \sum_{\mathbf{q}_\parallel} \tilde{S}(\mathbf{q_\parallel}) U_d^{\text{2D}}(\mathbf{q_\parallel}) \tilde{S}^\dagger(\mathbf{q_\parallel}) e^{i \mathbf{q}_\parallel \mathbf{r}_\parallel}   $
      \endtabular\\
      
      \psframebox[linearc=0.14,cornersize=absolute,framesep=8pt]{\small
	$\varepsilon_d^{\text{2D}}(\mathbf{q_\parallel})$
	$\quad \Rightarrow \quad U_d^{\text{2D}}(\mathbf{q_\parallel}) = \frac{v_d^{\text{3D}}(\mathbf{q_\parallel}, z=0)}{\varepsilon_d^{\text{2D}}(\mathbf{q_\parallel})}$
      }
      &
      
      &
      \psframebox[linearc=0.14,cornersize=absolute,framesep=8pt]{\small
	$U_{\tilde{\alpha} \tilde{\beta}}^\text{2D}(\mathbf{r}_\parallel)$
      }\\

    \end{psmatrix}
    
    \ncline{1,1}{2,1}\ncline{->}{2,1}{3,1}
    \ncline{1,1}{1,3}\ncline{1,3}{2,3}\ncline{->}{2,3}{3,3}
    
    \ncline{3,1}{4,1}\ncline{->}{4,1}{5,1}
    \ncline{3,3}{4,3}\ncline{->}{4,3}{5,3}
    
    \ncline{5,1}{6,1}\ncline{->}{6,1}{7,1}
    \ncline{5,3}{6,3}\ncline{->}{6,3}{7,3}

    \ncline{7,1}{7,2}\ncline{7,2}{6,2}\ncline{->}{6,2}{6,3}
    